\documentclass[final,2p,times,twocolumn]{elsarticle}

\usepackage{hyperref}
\usepackage[utf8x]{inputenc}
\usepackage{graphicx}

\journal{Journal of Environmental Management}

\begin{document}

\begin{frontmatter}

\title{Translating Ecological Integrity terms into operational language to inform societies\tnoteref{mytitlenote}\tnoteref{CC_license}}
\tnotetext[mytitlenote]{{Received 13 April 2018; Received in revised form 3 September 2018; Accepted 10 September 2018} \href{https://doi.org/10.1016/j.jenvman.2018.09.034} {doi.org/10.1016/j.jenvman.2018.09.034}}

\tnotetext[CC_license]{© 2018. This manuscript version is made available under the CC-BY-NC-ND 4.0 license \href{http://creativecommons.org/licenses/by-nc-nd/4.0/}{http://creativecommons.org/licenses/by-nc-nd/4.0/}}

\author[ICM]{Silvia de Juan\corref{mycorrespondingauthor}}
\cortext[mycorrespondingauthor]{Corresponding author: \href{mailto:sdejuanmohan@icloud.com}{sdejuanmohan@icloud.com}}

\author[NIWA]{Judi Hewitt}
\author[CCM]{Maria Dulce Subida}
\author[UoA]{Simon Thrush}

\address[ICM]{Institute of Marine Science (ICM-CSIC), Passeig Marítim de la Barceloneta, n° 37-49, 08003 Barcelona, Spain.}
\address[NIWA]{National Institute of Water and Atmospheric Research, P.O. Box 11-115, Hamilton, New Zealand.}
\address[CCM]{Núcleo Milenio – Center for Marine Conservation, Estación Costera de Investigaciones Marinas, Pontificia Universidad Católica de Chile, Santiago, Chile.}
\address[UoA]{Institute of Marine Sciences, The University of Auckland, Auckland, 1142, New Zealand.}
\begin{abstract}
It is crucial that societies are informed on the risks of impoverished ecosystem health for their well-being. For this purpose, Ecological Integrity (EI) is a useful concept that seeks to capture the complex nature of ecosystems and their interaction with social welfare. But the challenge remains to measure EI and translate scientific terminology into operational language to inform society. We propose an approach that simplifies marine ecosystem complexity by applying scientific knowledge to identify which components reflect the state or state change of ecosystems. It follows a bottom-up structure that identifies, based on expert knowledge, biological components related with past and present changing conditions. It is structured in 5 stages that interact in an adaptive way: stage 1, in situ observations suggest changes could be happening; stage 2 explores available data that represent EI; stage 3, experts’ workshops target the identification of the minimum set of variables needed to define EI, or the risk of losing EI; an optative stage 4, where deviance from EI, or risk of deviance, is statistically assessed; stage 5, findings are communicated to society. We demonstrate the framework effectiveness in three case studies, including a data poor situation, an area where lack of reference sites hampers the identification of historical changes, and an area where diffuse sources of stress make it difficult to identify simple relationships with of ecological responses. The future challenge is to operationalize the approach and trigger desirable society actions to strengthen a social-nature link.
\end{abstract}


\end{frontmatter}

\section{Introduction}

With an increasing demand for natural resources in a world of rapid biodiversity loss and environmental change, society needs to be well-informed about the consequences of changes in the environment (Cardinale et al., 2012). A two-way communication between scientists and society is critical, as there is an increasing demand to find new ways to conceptualise problems and find solutions in liaison with managers, politicians and common citizens (Carpenter et al., 2009; Castree, 2015; Chapin et al., 2010; Leslie and McLeod, 2007). We, ecologists, need to reduce the gap between the scientific knowledge we generate and its potential contribution to the well-being of societies. This social-ecological interaction is called the “new social contract” by the State of the Planet Declaration (2012), and, as Castree (2015) notes, we “need to link high quality, focussed scientific research to new policy-relevant interdisciplinary efforts for global sustainability”. But this “new social contract” implies that societies, encompassing decision-makers to end-users, need to be informed by credible and relevant evidence of not only the nature of environmental changes but on how those changes might feedback to affect the welfare of societies. In other words, Anthropocene societies need to be capable of not only building evidence-based relevant policy directed towards local-to-global problems, but also accurately measure the impact of those policies on the sustainability of human – natural systems.

\begin{figure*}
\centering
\includegraphics[width=0.9\textwidth]{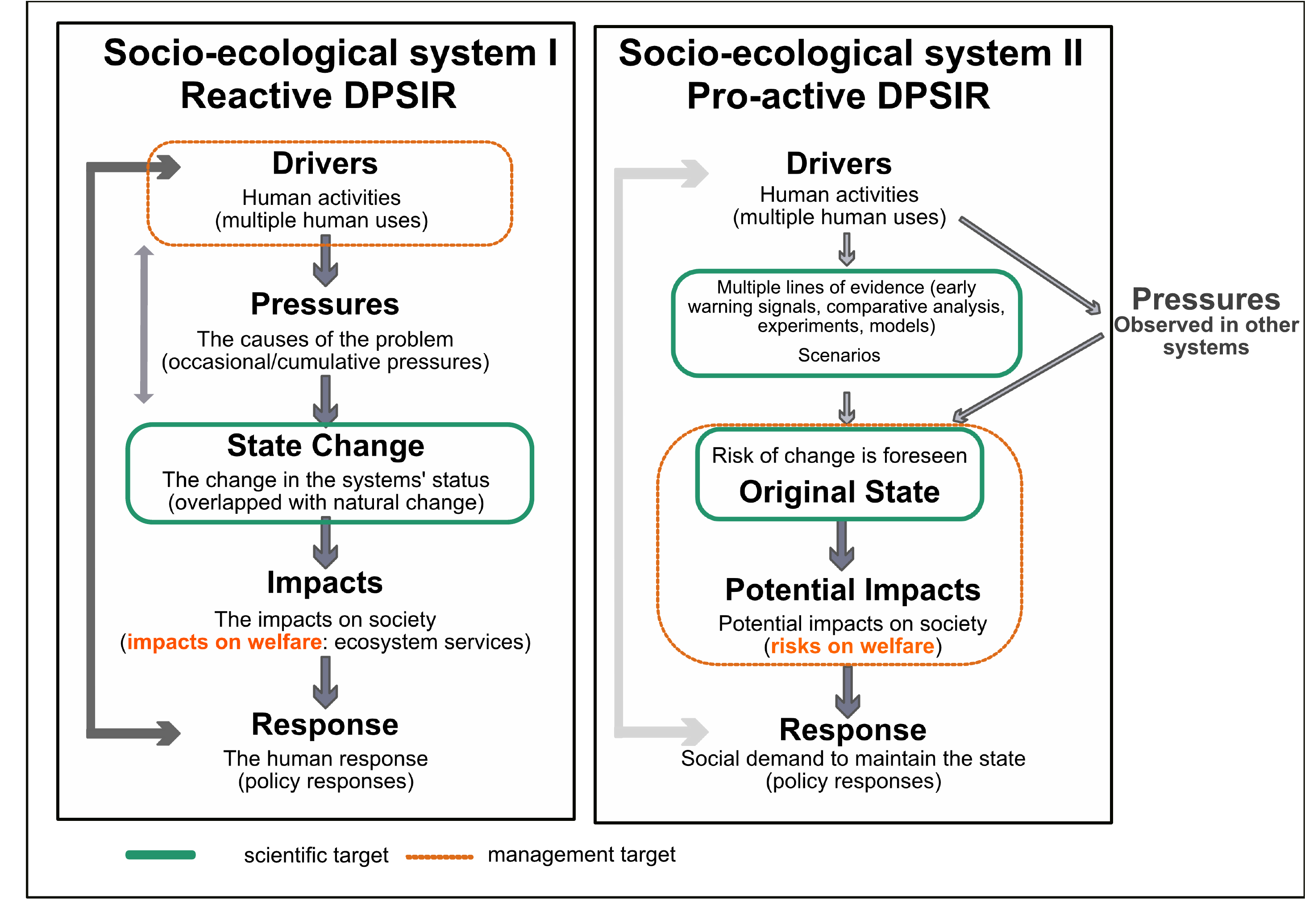}
\caption{The reactive and pro-active DPSIR framework: original framework adapted from Atkins et al. (2011) and Elliot et al. (2017).}
\label{fig:fig1}
\end{figure*}

The DPSIR framework (Drivers–Pressures–State change–Impact–Response) has been proposed as a systems-based approach that captures key relationships between society and the environment, and it is deemed useful for communicating environmental research to non-scientists (Atkins et al., 2011; Mangi et al., 2007). DPSIR seeks to integrate ecological and social information in a framework that takes account of the impacts of human activities on the functioning of ecosystems and the effects on society, and then introduces the need to apply measures to prevent or control adverse changes (Lonsdale et al., 2015). Moreover, it has been extended to incorporate different drivers and pressures, and interactions between these, to provide a nested framework to prioritise efforts and manage in real-world systems (Atkins et al., 2011; Elliott et al., 2017a). We might easily record information on the drivers and pressures, but we, as ecologists, are interested on the state change of ecosystems, either when changes are already observable or when these are foreseen (the desirable target of management). In order to reflect real-world scenarios of ecosystem change, we propose two DPSIR frameworks: a reactive DPSIR, aiming at finding solutions to existing impacts, and a pro-active DPSIR, aiming at forecasting potential pressures to find ways to minimize changes. The reactive framework is intended to (i) recover the original state, or (ii) reach an alternative sustainable state. In the pro-active framework (Fig.\ref{fig:fig1}) the response will come from within the initial state with the aim of maintaining it. Note that the framework includes several feedbacks between scientists, managers and society, as scientific findings need to inform management decisions, management actions should have an effect on the state change of ecosystems, and both managers and scientists should consolidate efforts to communicate with society. 

The DPSIR represents many scientific challenges but a major bottleneck to its effective implementation is to communicate this knowledge to promote responsible societies (i.e., Impact-Response components) (Elliott et al., 2017b). In order to address this challenge, we need to 1) apprise the ecosystem status in a variety of human disturbance scenarios, that encompass current status and alternative sustainable uses, reflecting real-world scenarios, and 2) identify integrative indicators of state/state change to transmit complex ecological concepts to society that lead to co-designed solutions.

A new socio-ecological perspective is particularly crucial in coastal areas, where there is a strong cultural and economic dependency of societies on marine ecosystems, and the increasing pressures of human activities might cut off the flow of benefits garnered from marine ecosystem services, jeopardising the long-term well-being of societies (de Juan et al., 2017; MEA, 2005). Considering the diverse nature of impacts on the marine environment, and even the more diverse range of ecosystem responses operating over multiple space and time scales, capturing this complexity by a common metric has been challenging (Kappel et al., 2009; Rombouts et al., 2013). Over the past decades, there has been a plethora of indicators of the ecosystem status, partly triggered by policies like the European Water Framework Directive (2000/60/EC; European Commision, 2000) and the Marine Strategy Framework Directive (2008/56/EC, European Commision, 2008). However, these indicators tend to deal with one major stressor or kind of response from the ecosystem and, in general, they have been designed for a scientific or decision-maker use rather than to inform society at large (e.g., Blanchet et al., 2008; Borja et al., 2008b; Pinto et al., 2009; Van Hoey et al., 2010). The Ocean Health Index (Halpern et al., 2012) is an integrative index that informs on global status of the ocean, but this metric does not consider cumulative or multiple stressor impacts at local or regional scales, and, therefore, operates at scales not relevant across society, but see the more recent OHI+ index designed to measure the ocean health at regional or local scales by independent groups of experts (http://www.oceanhealthindex.org/ohi-plus/portal). The welfare of society ultimately relies on a set of ecosystem structures and processes that are essential to maintain the system’s resilience and its ability to provide goods and services (Müller and Burkhard, 2007). Therefore, societies rely on functional ecosystems that are resilient to external pressures (Tett et al., 2013). The existing set of indicators of the ecosystem status generally illustrate the ecosystem structure, but provide little information on the ecosystem functioning (Borja et al., 2008a) and, therefore, on its capacity to provide services to society. 

A metric(s) that informs of the state or state change of marine ecosystems must encompass complex and scientifically sound information. Ecological Integrity (EI) has been proposed as a concept that captures the complex nature of ecosystems and its interaction with social welfare (Costanza et al., 1992; Karr, 1993). There are numerous definitions for EI but, in general terms, it is a holistic term that seeks to capture our sense of nature, its functionality and self-organising capacity (Tett et al., 2013). In fact, it is perhaps better understood by its absence rather than its presence. Thus, it depends on the wide-ranging perception of nature by societies. Despite the appropriateness of this concept for our objective, the challenge remains on finding how to translate EI terminology into operational language to inform decision makers and society at large. Our approach seeks to simplify complexity based on ecological knowledge, by applying this knowledge to identify which ecosystem components reflect EI. These components should be monitored to inform societies on the impact of existing or potential changes (and thus risk of losing EI), and ultimately aiming to trigger management responses.

\section{Indicators of Ecological Integrity: a bottom-up process }

While we need to operationally define EI and ensure the concept fits our purpose, the openness of the language does aid in the wide communication of complex and complicated concepts, similar to terms such as biodiversity or ecological sustainability. Elliott (2011) described “healthy” ecosystems as a property that protects against the “ecosystem pathologies” of Harding (1992), and monitoring “health” consists of “detecting when things go wrong” (note the similarity in meaning of Ecosystem Integrity, Ecosystem Health, Good Environmental Status). EI should be assessed through multiple metrics that quantify the multi-disciplinary essence of this concept and ultimately inform of the value of an area for society. There is a general agreement that utilizing a suite of indicators is the best approach to understanding ecosystem responses to drivers of change (Boldt et al., 2014). Importantly, the multiple metrics that represent EI must be related with the ecosystem state change produced by both occasional and cumulative disturbance, in order to be informative of real-world scenarios. Cumulative disturbance acts in space and in time (Thrush et al., 2013), which implies our metrics must also reflect these dimensions. From a set of components that represent EI, some might be individually correlated with a specific stressor but, overall, they should inform of the overall structure, function and resilience of the ecosystem, while being non-redundant and complementary (Boldt et al., 2014). 

Measures that represent EI should be knowledge-based and work in an adaptive way, by being responsive to continuous knowledge generation. The EI indicators should be somewhere between a large set of individual indicators, that are too confusing to illustrate the biggest picture, and aggregated indices that mask complexity (Hayes et al., 2015). In summary, the essential characteristics of EI indicators to accurately inform society are: i) sensitive to potential stressors; ii) effective over a range of space and time scales; iii) reflect the multi-component nature of ecosystems; iv) be justifiable by scientific knowledge; v) be adaptive to continuous knowledge generation; vi) be adaptive to changing environmental conditions; vii) be cost-efficient in time and expense; viii) be informative to society. Also, from an operational approach, we should try to avoid unreasonable data requirements (Borja et al., 2013). What we propose is a bottom-up process that identifies, based on current expert knowledge, biological components related with past and present changing conditions of the ecosystem.

Table S1 includes an example of the components of marine ecosystems that could be monitored to jointly inform of EI. The proposed variables would range from a broad spectrum to a minimum set and would be monitored within a hierarchical framework. This structure allows incorporating ecosystem processes and features that operate at different scales. In summary, these indicators are: 1) variables related to biological communities’ composition and range from broad- to local-scale; 2) variables related to things changing on a temporal or a spatial scale (or both), e.g., temporal as observed changes in time in an area, and spatial as observed changes between environmentally similar areas; 3) variables that capture the essence of EI, and indicate its loss due to stress, in a combined way that is easy to understand and justifiable. 

\subsection{Assessment framework}
\begin{figure*}
\centering
\includegraphics[width=0.9\textwidth]{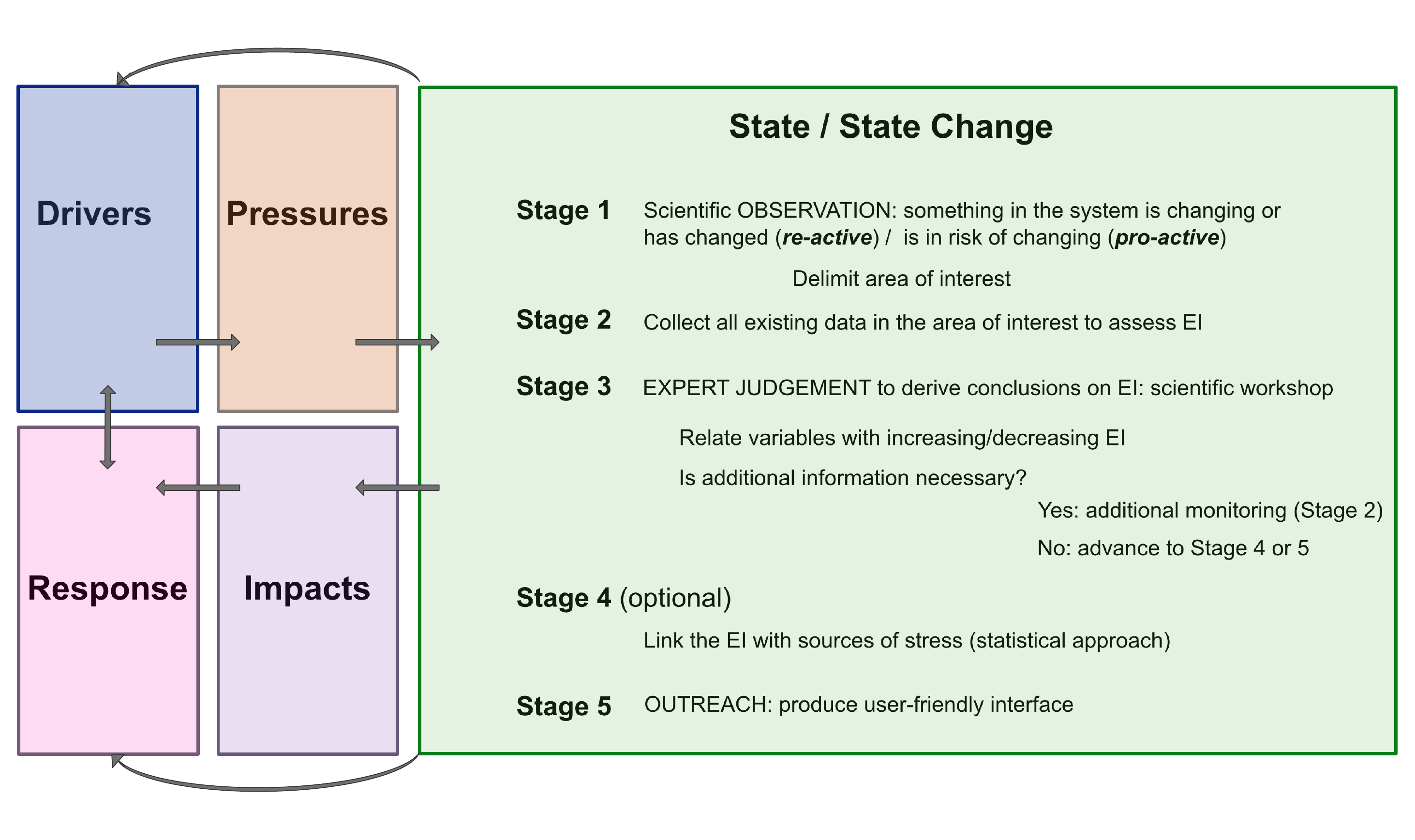}
\caption{Components of the proposed bottom-up approach to assess Ecological Integrity of an area. The approach also considers feedbacks between scientists, managers and society.}
\label{fig:fig2}
\end{figure*}

We propose a bottom-up process within a hierarchical framework that has the following steps: 1) observation; 2) expert judgement; 3) outreach. These steps are achieved in stages: A) collect available ecological data (Table S1); B) conduct an experts’ workshop to relate these data with EI; C) feedback process by increasing monitoring efforts if necessary; D) scientific assessment to produce a multi-metric index related with EI, or the risk of losing EI, in the area of interest (Fig.\ref{fig:fig2}). Large-scale and long-term data sets, that are crucial to monitor ecosystem changes, are rare in marine science mainly due to the cost (Hayes et al., 2015). Therefore, we believe that part of monitoring for EI should include triggers and responses when in situ scientific observation leads to conclusions of “something is wrong”. This structure has analogies with a System Thinking, with an essential feedback component that relates the different levels of the system: 1) observable changes, 2) patterns in changes, 3) complex interactions in the system that conditions its resilience, and 4) individual perceptions of the system that collectively would trigger responses  (Maani and Cavana, 2007). 

Our strategy begins with the reasons behind the desire to calculate EI in a particular place: have changes been observed or is there a need to simply catalog state? Alternatively, has knowledge on the existence of drivers of change in an area suggested that an ecosystem state change is likely? The first stage of the strategy (\textbf{Stage 1}) thus consists of stating the rationale and delimiting the area of interest by defining boundaries based on geophysical (e.g. estuary), human activities (e.g., fishing ground), or ecological components (e.g., seagrass habitat) (Fig.\ref{fig:fig2}).

The second stage (\textbf{Stage 2}) is to identify existing data in the area of interest that could be used to calculate variables representing EI. These can range from broad-scale to site-specific, which subsequently might need to be extrapolated to the area of interest (e.g., Table S1). We suggest a focus on a specific set of variables, adapted to the location and issue of interest, with a weighting of the different metrics that could be adjusted when disturbance regimes change, management and societal needs vary, or scientists warn of approaching thresholds. In addition, when adopting a pro-active DPSIR framework, we recommend deciding on the potential drivers of pressures with a focus on those that are predicted to change and the likely magnitude of those changes. The following step would be to gather historical data relating each of the variables in the registry with pressures that led or were envisaged to lead to a state change in other systems. The extrapolation to our system of interest of causal relationships between drivers (actually observed in our system) and pressures (observed in other systems), is expected to facilitate anticipating the impacts of a state change. Also, the identification of early warnings of change (Scheffer et al., 2009) and modelling potential scenarios of change are crucial scientific contributions in the pro-active context. 

The data might be obtained under different contexts that could involve regular monitoring generally conducted by environmental or management agencies (e.g., fishing landings or nutrient concentrations), fieldwork within research programs (e.g., diversity estimates or species densities), or citizen science (e.g., detection of jellyfish blooms). We claim special emphasis on available data sets that should not be ignored, as many regular monitoring programs, including fishery surveys, registers of oceanographic parameters, or even citizens recording species presence, have left an enormous data legacy (Edgar et al., 2016) that, particularly in data-poor regions, contains crucial information regarding the past status of marine life and its interactions with people and the climate (Roelfsema et al., 2016; Roemmich et al., 2012). Additionally, in the absence of conventional data, historical ecology approaches such as fishers knowledge, historical narratives and archives are valuable alternative approaches to understanding ecological dynamics and recovery potential (McClenachan et al., 2012; Kittinger et al., 2015); however, local knowledge needs to be treated with care as it can be biased due to end-users’ interests (Ruano-Chamorro et al., 2017). 

The third stage (\textbf{Stage 3}) involves the use of expert judgement to identify the minimum set of variables necessary to derive conclusions on EI of the area. This would include using general scientific knowledge to establish the key features and relevant variables for each feature in the area of interest and then identifying the relationships between the relevant variables with EI. This would also include the extrapolation to our system of interest of causal relationships between drivers (actually observed in our system) and pressures (observed in other systems). At this stage it would also be appropriate for the experts to consider how to relate each variable with a state that reflects a certain degree of increasing/decreasing EI and if necessary to assign the weights to each variable (e.g., adopting a fuzzy modelling approach; Game et al., 2017). Participatory approaches are useful to identify, and potentially score, the relevant variables that are related with the system’s change (Game et al., 2017). For example, Multi-Criteria Decision Analysis could be adopted to produce a multi-metric indicator in a transparent and objective way (Estévez and Gelcich, 2015). This process should be one of continuous knowledge generation, where decisions on the need to collect additional data, from several or specific metrics, should be decided. This structure would contribute to working with an adaptive protocol for the assessment of EI. 

Sometimes the process would stop at this stage, by describing the state based on the obtained multi-metric indicator. However, often scientists and managers require information on change or risk of change in state in relation to existing pressures. In this case (\textbf{Stage 4}), statistical analyses or risk assessments of change are required. Statistical analysis could be done on the overall assessment of EI. However, analysis on individual components is helpful so that scientists and managers understand which particular elements of integrity are driving the observed changes (Martínez-Crego et al., 2010). Boldt et al. (2014) suggests that data-based approaches are not practical for assessing the effects of multiple stressors on multiple ecosystem components, rather that experts’ opinions and model-based simulations are the best tool. But model-based simulations assume the ability to develop a suite of appropriate models that address possible cumulative and multiple stressor effects, these are not currently available. Conversely, regressive data-driven techniques can provide important insights. For example, Thrush et al. (2008b) demonstrate how heavy metal toxicity interacting with sediment grain size change species distributions and Thrush et al. (2012) show how interaction networks can change across thresholds in sediment-benthic chlorophyll relationships. Importantly, the number of variables for which changes are detected could become a significant part of the assessment of EI. 

A decreasing EI, or evidence on risks to EI, is the scientific link between society and the drivers of use that ultimately could be modified by societal actions. This may represent a direct action to lower disturbance or stress levels (e.g., dredging disposals, fishing, pollutants discharges) or an indirect action in response to a driver that cannot be directly modified in the short-to-medium term (e.g., climate change). Therefore, at the bottom of the hierarchical structure, the multi-metric index is related with the “risk” that something is changing, the probability that some human activities are having impacts that might threaten the EI of an area and, ultimately, the well-being of society. \textbf{Stage 5}, the final and crucial step, is thus for science to integrate the EI indicators in a user-friendly interface to inform society, including managers, as to what is happening and why (e.g., Muntadas et al., 2017). The output must be both practical, easy to interpret and meaningful, for example, a weight of evidence against criteria or about time. Ultimately, the information on ecosystem components could be built into a categorical matrix, similar to a decision tree that relates changes in state with potential stressors and the risk of losing EI. Following this structure, the effects of specific stressors could be tracked down the matrix. These examples are a useful tool to communicate findings to stakeholders or to small groups where scientists can explain the outcome. However, outreach should also be focused on scientific dissemination initiatives, like scientific cafes, activities in museums or aquaria, or in association with artists to explore alternative communication channels, being crucial the use of plain and attractive language. In both cases, new communication and information technologies are currently a fundamental ally in the process of making complex scientific data more accessible for the society at large. Scientists should also make use of the social media, often full of misinformation, so scientific-based information gains more visibility for a large and diverse public. Regardless of the method adopted, for an effective outreach we need to establish a two-way communication with society, so their perceptions and beliefs nourishes our research goals (Varner, 2014).

\section{Indicators of Ecological Integrity in the real world}

It has been largely argued that the interpretation of ecosystem indicators requires the assessment of the obtained values against a reference, traditionally known by ecologists as baseline conditions (Duarte et al., 2008). However, there are two major drawbacks to baselines or reference states in the real-world of marine ecosystems. On one hand, this perspective is often biased due to the historical use, where the conditions assumed as reference in the present might in fact be linked to chronically modified ecosystems, since an individual might perceive as “natural” an already degraded ecosystem (Dayton et al., 1998; Plumeridge and Roberts, 2017). On the other hand, true reference values in naturally disturbed systems, like estuaries for instance, may be considered as indicating a degraded condition (Van Hoey et al., 2010) even in the absence of anthropogenic pressures. In fact, the dynamic nature of ecosystems implies that the state change due to human impacts overlaps natural dynamics, which in turn affect future states (Ladle and Gillson, 2008; Paoli et al., 2016). This drawback is more acute when historical knowledge of the system rarely exists (de Juan et al., 2018; Elliott et al., 2015). EI implies self-regulation of the system and, similarly to the concept of resilience, it is based on science and societal values. For the conservation of socio-ecological systems, our target is to define an agreed level of disturbance to which the system is resilient (i.e., identify potential switch points to alternative stable states, \textit{sensu} Scheffer et al., 2001), considering resilience as the insurance against the loss of valued ecosystem functions and services (Thrush et al., 2009). Therefore, instead of relying on ecosystem benchmarks and baselines, the proposed approach is founded on capturing the dynamics of ecosystem change that vary within resilient levels (similar to Tett et al., 2013), understood as the capacity of a system to maintain its structure, functions and feedbacks while being subjected, or in risk of being subjected, to change due to disturbance (Folke et al., 2004). Another relevant issue is the “desired state” that can vary amongst stakeholders, and it also might be influenced by what people perceive as nature (Levin and Möllmann, 2014). In this context, EI could be confronted to a set of environmental future scenarios defined with the feedback of stakeholders and society. In summary, conservation might be less about bridging the gap between scientific knowledge and societal perceptions and more on effectively explaining conservation targets to society (Stage 5 of the framework). Our proposed approach does not exclude the potential that some specific metrics could be assessed against specific benchmarks. After all, the essence of the approach is to be practical while evading rigid schemes to promote societal consciousness and involvement in conservation, to avoid a society that ignores sliding baselines and thresholds of change in marine ecosystems.

\section{Case studies}

\begin{figure*}
\centering
\includegraphics[width=0.9\textwidth]{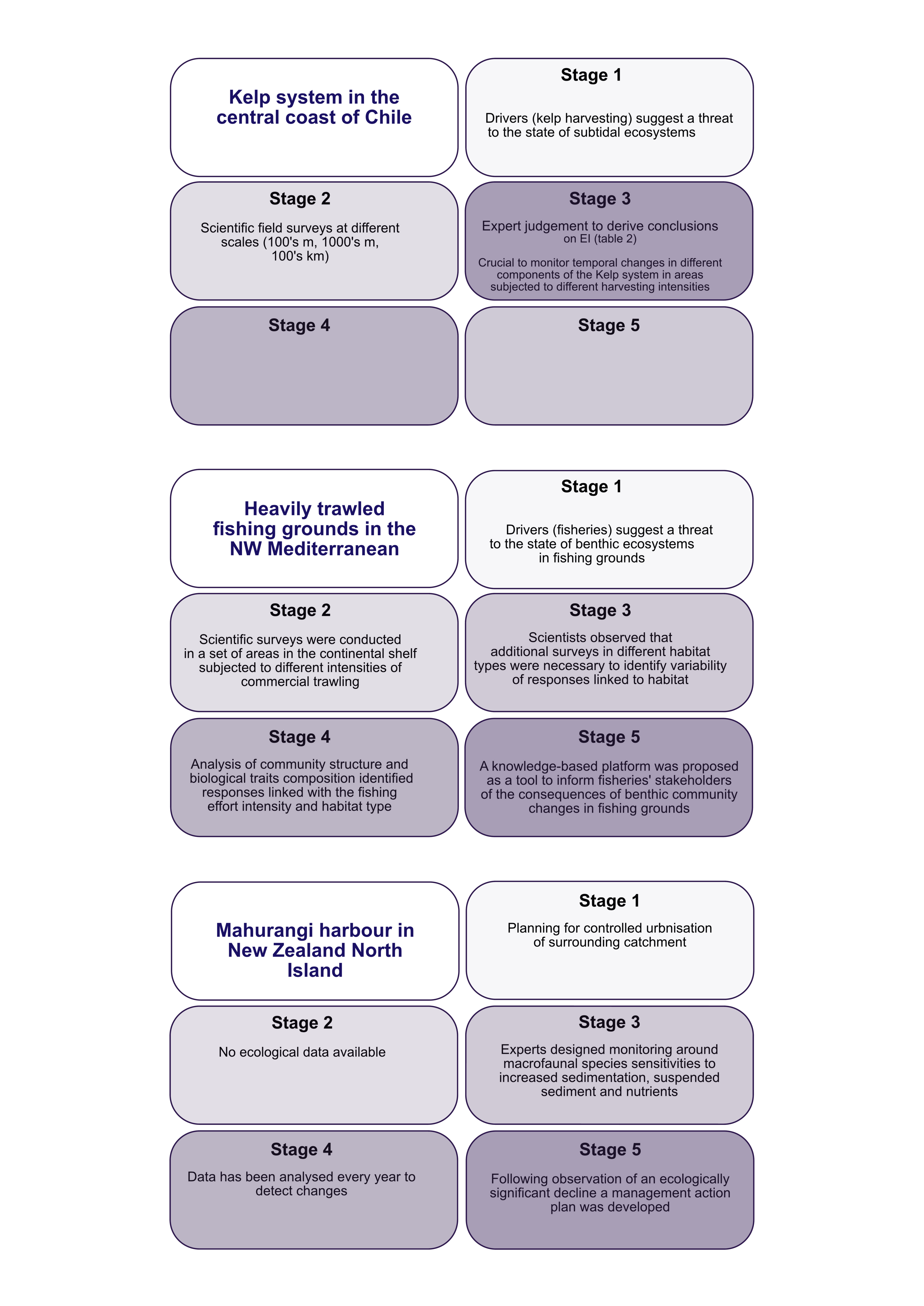}
\caption{Example of the applicability of the EI assessment framework to the case study of a) kelp harvesting in the central coast of Chile; b) benthic communities disturbed by trawling activities in the NW Mediterranean; c) Mahurangi harbour in New Zealand North Island.}
\label{fig:fig3}
\end{figure*}

Three different case studies, reflecting different coastal and maritime regions, environmental contexts and habitats, are described to illustrate the principles of the approach in real-world scenarios. These studies, conducted prior to the proposal of this approach, do not totally match the five Stages; however, these examples, including a summary of existing data (Table \ref{table:T1}), help to illustrate how our approach is sufficiently flexible to work with a diversity of scenarios and scientific studies. The first two case studies illustrate a reactive DPSIR, where we aim to assess the state change of the system, whereas the third case study illustrates a pro-active DPSIR, where we suspect there is a risk of losing the EI of the area (Fig.\ref{fig:fig3}).

\begin{table*}[]
\centering
\caption{Compilation of the variables from all case studies}
\label{table:T1}
\resizebox{\textwidth}{!}{%
\begin{tabular}{lll}
\hline
\textbf{CASE STUDY} & \textbf{Scale} & \textbf{Indicator} \\ \hline
\textbf{Central Chile kelp system} & Site & Density of habitat forming species \\
 &  & Changes in key-species abundance \\
 &  & Trophic structure \\ \hline
\textbf{NW Mediterranean fishing ground} & Site & Density of habitat forming species \\
 &  & Modification of seabed \\
 &  & Changes in key-species abundance \\
 &  & Biological traits composition \\
 &  & Functional redundancy \\ \hline
\textbf{Mahurangi harbour in New Zealand North Island} & Site & Sediment characteristics \\
 &  & Microphytobenthos biomass \\
 &  & Key infauna and epifauna \\
 &  & Infaunal community structure and biodiversity \\ \cline{2-3} 
 & Estuary & Spatial gradients \\
 &  & Coverage of Atrina beds \\
 &  & Infaunal species alpha and beta diversity \\ \hline
\end{tabular}%
}
\end{table*}

\subsection{Kelp system in the central coast of Chile}

\textbf{Stage 1}: Commercial harvesting of habitat structuring species, like kelps, is likely to produce pervasive impacts on the ecosystem they create (Mac Monagail et al., 2017). In Chile, more than 250.000 tons of the subtidal kelp \textit{Lessonia trabeculata} and its intertidal congeneric \textit{L. nigrescens} are harvested each year; this activity is highest in northern regions but is increasing southwards (Vásquez et al., 2012). Despite this, the magnitude of the direct and indirect effects kelp harvesting may have on kelp-associated communities, many of them of commercial importance, is largely ignored for subtidal kelp forests in Chile (de Juan et al., 2018; Vásquez et al., 2012). Small-scale fisheries in the area, including kelp harvesting, are subjected to two management regimes: Territorial Users Rights for Fisheries (TURFs) and open access areas (OAAs) that create a mosaic of fishing access regimes in the coastal landscape.

\textbf{Stage 2}: In intertidal kelp systems of northern Chile, removal of high numbers of adult sporophytes in small areas has been associated with increased access of grazers to harvested kelp grounds (Santelices and Ojeda, 1984; Vásquez and Santelices, 1990). Manipulative experiments in subtidal kelp grounds of central Chile showed that grazers are capable of reducing kelp growth and increase blade tissue loss; a broad-scale (100’s km) field survey in the area revealed this effect may be stronger in OAAs with no upwelling influence (Pérez-Matus et al., 2017a). Results of a single site snapshot study where kelp patches in different stages of post-harvesting recovery were surveyed (scale of 100’s of m), revealed an increase in the amount of grazer snails in harvested patches in relation to non-harvested ones (Pereira et al., 2015). When considering different fishing management regimes (1000’s of m), grazer invertebrates, including snails, responded to kelp harvesting with increased abundance but only in OAAs. Adult kelps, in turn, were larger and more abundant inside TURFs (Subida et al. in preparation). 

\textbf{Stage 3}: Variability in kelp-associated communities may be explained by spatial effects taking place at multiple scales (Lamy et al., 2018). Data from the area of interest suggest that the variable offering the best picture of the community-wide effects of kelp harvesting is the abundance of grazer snails. The direction and magnitude of this effect is expected to depend on broad-scale oceanographic processes (upwelling) and fisheries management regimes operating at the small-scale: harvested kelp grounds in low upwelling OAAs are expected to be more vulnerable to grazer outbreaks. Since fishing effects are expected to cascade throughout the entire food web in this system (Pérez-Matus et al., 2017b), the increased abundance of grazer snails in harvested kelp grounds of OAAs may occur because: i) in OAAs there is a higher fishing effort on potential consumers of \textit{Tegula tridentata}, reducing the snail mortality by predation (Pérez-Matus et al., 2017a); and/or (ii) inside TURFs controlled harvesting effort limits the number of adult sporophytes removed per unit area, limiting the access of grazing snails to the harvested kelp ground (Konar and Estes, 2003; Pereira et al., 2015; Vásquez and Santelices, 1990).

At the current (and limited) state of knowledge, assessing the change in state in this ecosystem (and therefore, quantify deviance from EI; Stage 4) requires setting adaptive benchmark areas against which the abundance of grazer snails can be compared. Benchmark areas must account for (i) the fisheries management regimes in force, (ii) varying levels of harvesting intensity, and (iii) small-scale upwelling processes. However, considering the habitat forming feature of the exploited kelp resource, benchmarking and monitoring of harvesting effects should also consider temporal changes at the wide-community level.

\subsection{Benthic communities from the heavily trawled NW Mediterranean}
\textbf{Stage 1}: The negative effects of trawling activities on the seabed have been evidenced through numerous studies (e.g., Auster et al., 1996; Jennings et al., 2005; Kaiser et al., 2006), however, the links between the benthic community structure, its functioning and provision of services, and trawling activities are not always easy to establish, particularly in chronically trawled areas. This is the case of the NW Mediterranean, where historical trawling has left the seabed with no real reference sites (de Juan et al., 2009). 

\textbf{Stage 2}: Scientific surveys were conducted to collect data on benthic communities from fishing grounds in the NW Mediterranean, and where different levels of trawling effort were detected. 

\textbf{Stage 3}: Despite the absence of reference sites, and thus of knowledge on the structure of non-trawled benthic communities, patterns of community responses to fishing intensity were identified and interpreted with support of the sound knowledge scientists have acquired on the structure and function of soft-bottom marine ecosystems (de Juan et al., 2007). The analysis of data collected in areas subjected to variable effort matched our predictions on the response of benthic communities to trawling activities (\textbf{Stage 4}): higher densities of sessile and emergent suspension feeders in the less trawled areas, compared to higher densities of small, burrowing, deposit feeders in the heavily trawled sites (de Juan and Demestre, 2012). Also, environmental modifications were observed in the most heavily trawled sites, with a homogenised muddy bottom and increased near-bottom turbidity (Palanques et al., 2001). 

Benthic community responses are influenced by environmental context and, in the NW Mediterranean, responses proved to be tightly linked with habitat type. The survey of thirteen additional sites having different sediment composition and variable trawling disturbance levels (back to \textbf{Stage 2}) allowed the comparison of benthic community responses to trawling within habitat types (de Juan et al., 2013). Response patterns depended on the sampled community (\textbf{Stage 4}): a heavily trawled rhodolith bed harboured significantly higher densities of vulnerable fauna and higher species richness than less trawled muddy bottoms. However, when comparisons were restricted to similar habitat types, observations on the benthic community response patterns confirmed expectations, i.e. less vulnerable fauna and lower richness linked with increasing effort, even in rhodolith beds (de Juan et al., 2013). These results were not totally unexpected as rhodolith beds are known to harbour high species diversity (Barbera et al., 2012) and this implies a different baseline from which to assess benthic community changes due to trawling activities. In this case, data interpretation based on experts’ knowledge allowed to arrange the trawled sites in a degree of deviance from EI.

\textbf{Stage 5}: studies conducted in this area have established a solid baseline for the use of Biological Traits of benthic communities as an “easily interpreted” indicator of the ecosystem status in trawling grounds that could rate an area within a poor-good status framework following MFSD approaches (e.g., de Juan and Demestre, 2012). Based on this approach, Muntadas et al. (2017) introduced a knowledge-based platform as a proposal for a decision support tool for fisheries’ stakeholders, and also to raise societal consciousness on the state of ecosystems in trawling grounds. Additionally, observed responses to trawling activities have been related with alterations in the provision of services by benthic communities, particularly of their role in providing habitat (feeding and reproduction grounds) for commercial species (Muntadas et al., 2015; 2014). The challenge now is to effectively inform society on the impacts these activities will have in their long-term well-being.  

\subsection{Mahurangi harbour in New Zealand North Island}
\textbf{Stage 1}: Terrestrial sediment entering estuaries at rates above natural levels due to changes in land use have been identified as a major stressor on estuaries and harbours (Ellis et al., 2000; GESAMP, 1994; Gray, 1977). In the early 1990s Mahurangi catchment was identified as an area with a high potential for change in land use.  

\textbf{Stage 2}: No ecological data was available for the harbour.

\textbf{Stage 3}: There was a long history of ecological research suggesting that sediment type was a major driver of the structure and function of benthic macrofaunal communities (Gray, 1977; Snelgrove and Butman, 1995), although at this stage there were few cause and effect studies relating species occurring in the estuary to particular sediment characteristics. The harbour had extensive intertidal sand and mud flats, very little of which was vegetated, and deep high-flow tidal channels which supported extensive beds of the suspension-feeding bivalve \textit{Atrina zelandica}. The dense beds of \textit{Atrina} were observed to support sponges, other epifauna, small fish and infauna (Cummings et al., 1998). A monitoring programme was developed (Cummings et al., 1995) that focussed on a selection of intertidal and subtidal macrofaunal species likely to have differing sensitivities to different types of stressors (sedimentation, suspended sediment and nutrients). Abundances of these species were monitored at 5 intertidal and three subtidal sites arrayed along a gradient of mud content and observed turbidity. At the subtidal sites, growth and condition of \textit{Atrina} were also monitored.  Monitoring was conducted twice per year at the subtidal sites and quarterly at the intertidal sites.  Annually, all species (not just the selected species) were identified and enumerated.  During this time period a number of experimental additions of terrestrial sediments to intertidal sand and mud flats were conducted (Ellis et al., 2002; Hewitt et al., 2003; Norkko et al., 2002; Thrush et al., 2003).

\textbf{Stage 4}:  After 5 years, trends in abundance of each species and changes in community structure and composition at each site, as well as trends in the growth rate and condition of \textit{Atrina} were statistically assessed on a bi-annual basis.  

\textbf{Stage 5}:  Data was summarised as the number of species that were predicted to be sensitive to sedimentation and suspended sediment that showed trends in the predicted direction and reported to the management agency. In 2003 an ecologically significant decline in the condition of certain biota was reported (Cummings et al., 2003). An ecological assessment concluded that there were (i) estuary-wide declines in the abundance of some sedimentation-intolerant taxa, and (ii) general increases in the abundance of other groups, and that (iii) these changes are consistent with a model of large scale increases in sedimentation and benthic resuspension across the estuary. A Mahurangi Estuary Management Plan was established in 2004.  

Over the years there have been iterations between Stages 2 - 5 as more information has become available on the effects of terrestrial sediment on estuarine ecosystems. Habitat fragmentation and homogenisation has been identified as the large beds of \textit{Atrina} became smaller patches and finally disappeared (Halliday and Cummings, 2011). Beta diversity at a number of scales has been estimated (de Juan and Hewitt, 2013), and factors affecting recovery (and resilience) have been investigated experimentally (Lohrer et al., 2006; Thrush et al., 2008a). Estimates of health associated with mud content and metal contaminants have been developed using multivariate models (Hewitt et al., 2005) and a summary measure of functional health based on redundancy of biological traits has also been developed (Rodil et al., 2013). An important management step was taken in 2008, when managers and industry stakeholders from the surrounding catchment took part in the development of a Bayes net model to identify issues and likely solutions (Hume et al., 2009). Routine reporting has expanded to include changes in sediment characteristics and three summary health measures, as well as numbers of trends likely to be a result of changing sediment inputs and report cards are issued yearly. However, to date none of the measures have been integrated into a single measure of EI. More weight is given to observed trends in key species, total number of species changes and the health indices, than changes in minor species and sediment characteristics (Halliday and Cummings, 2011).

\section{Conclusion}
In order to ensure a long-term sustainable development, societies need to be informed of the “health” or “environmental status” of ecosystems. In this context, EI is particularly useful as it is a holistic concept that integrates the perception of nature by societies and the functionality of ecosystems. By adopting this concept, we propose a framework to operationalise EI in real-world scenarios to inform decision makers and society at large on the state, or state change, of ecosystems. We illustrate our proposed approach with a DPSIR framework, by distinguishing two forms of the framework: a reactive and a pro-active DPSIR, that vary with regards to the existence of pressures or to suspects on the risks of pressures. Our bottom-up approach is initiated with scientists in the field either observing “something is wrong” or “there is a risk of disturbance”, that would trigger a process that aims to monitor potential changes in the system from a biological perspective. The objective of the process is an ecologist interpretation of a system’s state structured in five principal stages. After the stage 1, when in situ observations suggest changes might/could be happening, stage 2 consists on exploring the available data that could represent the EI concept. Then, stage 3, aims to take advantage of the experts’ knowledge to identify the minimum set of variables needed to define EI, or the risk of losing EI, in the area of interest; importantly, this process (stages 2-3) should follow an adaptive framework nourishing form continuous knowledge generation. Depending on data availability, the framework could move forward to a stage 4, where deviance from EI, or risk of deviance, is identified though statistical techniques adapted to multivariate data frames. Finally, a crucial stage 5 targets the communication of findings to society through a two-way communication channel, aiming to achieve a common understanding of scientific findings. The lack of systems thinking could reduce practical credibility of the approach to scientific community, as well for society. However, the five steps’ structure also targets the understanding of dynamic feedbacks in complex systems (Nguyen and Bosch, 2012). The first level of our approach detects the symptoms of the problem and subsequently focus on the collection of available data that could lead to the second level of the system, identifying patterns of change in time or space. The complexity of ecosystem components’ interactions is in the essence of the EI concept, relying on the resilience and self-organizing capacity of the system, which is assessed in experts’ workshops that should nourish from a feedback structure within scientists and with society (Fig.\ref{fig:fig2}). Importantly, the approach effectiveness was tested in a diverse set of scenarios that illustrated the reactive and pro-active socio-ecological frameworks, and provided positive outcomes encouraging a more generalised adoption of the approach. Future challenges to operationalise this approach mainly rely on identifying how society is expected to use this information once the state or state change of the system has been communicated and linked with potential loss of social well-being. We acknowledge our proposal is the baseline for a broader and more complicated adaptive management process, as it is essential that there is a clear link between the reporting of EI of an area and society actions (e.g. stopping an activity, demanding more Marine Spatial Planning, improving land management, restoration, etc.), therefore, the link between change and reported metric(s) is critical. If the EI indicator could be decomposed in variables linked to specific stressors, the monitoring of integrity could be focused on a set of indicators in specific times, when the probability of a stressor, or set of stressors, is higher. This property would help to understand the implications of change in any index, so it could be linked to changes in EI in a transparent way. Additionally, and considering that there is a time-lag between the assessment of ecosystem status and the management activities, we stress the importance of dealing with a set of indicators that forewarn society, allowing time to demand for actions and reduce stress sources to avoid thresholds of change (being this of particular importance on a pro-active case study). Overall, the assessment of the integrity of socio-ecological systems at regional scales should both inform societies and support informed decision making while acting as a consolidated early warning system. 

\section*{Acknowledgements}
We thank two anonymous reviewers for comments that helped to improve the manuscript. Funding for JH was provided by NIWA Coasts and Oceans Centre (MBIE CORE). SdJ was funded by H2020-Marie Skłodowska-Curie Action MSCA-IF-2016 [Project ID: 743545]. MDS was funded by FONDECYT 1130580 grant and ICM grant CCM – RC130004 from Ministerio de Economía, Fomento y Turismo (Chilean Government).

\section*{Appendix A. Supplementary data}
\label{SI:S1}
Supplementary data to this article can be found online at: 
\href{https:// doi.org/10.1016/j.jenvman.2018.09.034}{https:// doi.org/10.1016/j.jenvman.2018.09.034}.

\section*{References}
Atkins, J.P., Burdon, D., Elliott, M., Gregory, A.J., 2011. Management of the marine environment: integrating ecosystem services and social benefits with the DPSIR framework in a systems approach. Mar. Pollut. Bull. 62, 215–226. doi:10.1016/j.marpolbul.2010.12.012

Auster, P., Malatesta, R.J., Langton, R.W., Watling, L., Valentine, P.C., Donaldson, C.S., Langton, E.W., Shepard, A.N., Babb, I.G., 1996. The impacts of mobile fishing gear on seafloor habitats in the Gulf of Maine (Northwest Atlantic): implications for conservation of fish populations. Rev. Fish. Sci. 4, 185–202.

Barbera, C., Moranta, J., Ordines, F., Ramon, M., Mesa, A., Díaz-Valdés, M., Grau, A.M., Massuti, E., 2012. Biodiversity and habitat mapping of Menorca Channel (western Mediterranean): implications for conservation. Biodivers. Conserv. 21, 701–728. doi:10.1007/s10531-011-0210-1

Blanchet, H., Lavesque, N., Ruellet, T., Dauvin, J.C., Sauriau, P., Desroy, N., Desclaux, C., Leconte, M., Bachelet, G., Janson, A., 2008. Use of biotic indices in semi-enclosed coastal ecosystems and transitional waters habitats—Implications for the implementation of the European Water Framework Directive. Ecol. Ind. 8, 360–372. doi:10.1016/j.ecolind.2007.04.003

Boldt, J., Martone, R., Samhouri, J., Perry, I., Itoh, S., Chung, I.K., Takahashi, M., Yoshi, N., 2014. Developing Ecosystem Indicators for Responses to Multiple Stressors. Oceanography 27, 116–133. doi:10.5670/oceanog.2014.91

Borja, A., Bricker, S.B., Dauer, D.M., Demetriades, N.T., Ferreira, J.G., Forbes, A.T., Hutchings, P., Jia, X., Kenchington, R.A., Carlos Marques, J., Zhu, C., 2008a. Overview of integrative tools and methods in assessing ecological integrity in estuarine and coastal systems worldwide. Mar. Pollut. Bull. 56, 1519–1537. doi:10.1016/j.marpolbul.2008.07.005

Borja, A., Dauer, D.M., Diaz, R.J., Llanso, R., Muxika, I., Rodriguez, J., Schaffner, L., 2008b. Assessing estuarine benthic quality conditions in Chesapeake Bay: A comparison of three indices. Ecol. Ind. 8, 395–403. doi:10.1016/j.ecolind.2007.05.003

Borja, A., Elliott, M., Andersen, J.H., Cardoso, A.C., Carstensen, J., Ferreira, J.G., Heiskanen, A.-S., Marques, J.C., Neto, J.M., Teixeira, H., Uusitalo, L., Uyarra, M.C., Zampoukas, N., 2013. Good Environmental Status of marine ecosystems: What is it and how do we know when we have attained it? Mar. Pollut. Bull. 76, 16–27. doi:10.1016/j.marpolbul.2013.08.042

Cardinale, B.J., Duffy, J.E., Gonzalez, A., Hooper, D.U., Perrings, C., Venail, P., Narwani, A., Mace, G.M., Tilman, D., Wardle, D., Kinzig, A.P., Daily, G.C., Loreau, M., Grace, J.B., Larigauderie, A., Srivastava, D.S., Naeem, S., 2012. Biodiversity loss and its impact on humanity. Nature 486, 59–67. doi:10.1038/nature11148

Carpenter, S.R., Mooney, H.A., Agard, J., Capistrano, D., DeFries, R.S., Díaz, S., Dietz, T., Duraiappah, A.K., Oteng-Yeboah, A., Pereira, H.M., Perrings, C., Reid, W.V., Sarukhan, J., Scholes, R.J., Whyte, A., 2009. Science for managing ecosystem services: Beyond the Millennium Ecosystem Assessment. PNAS 106, 1305–1312. doi:10.1073/pnas.0808772106

Castree, N., 2015. Geographers and the Discourse of an E arth Transformed: Influencing the Intellectual Weather or Changing the Intellectual Climate?. Geog. Res. 53(3), 244-254.

Chapin, F.S., Carpenter, S.R., Kofinas, G.P., Folke, C., Abel, N., Clark, W.C., Olsson, P., Smith, D.M.S., Walker, B., Young, O.R., Berkes, F., Biggs, R., Grove, J.M., Naylor, R.L., Pinkerton, E., Steffen, W., Swanson, F.J., 2010. Ecosystem stewardship: sustainability strategies for a rapidly changing planet. Trends Ecol. Evol. 25, 241–249. doi:10.1016/j.tree.2009.10.008

Costanza, R., Norton, B., Haskell, B., 1992. Ecosystem Health. New goals for environmental management. Island Press, Washington D.C.

Cummings, V.J., Hewitt, J.E., Wilkinson, M.R., Thrush, S.F., Turner, S.J., 1995. Mahurangi Estuary biological monitoring programme - report on data collected during the first year of monitoring (No. Report No. ARC60207). NIWA Consultancy.

Cummings, V.J., Nichols, P., Thrush, S.F., 2003. Mahurangi Estuary ecological monitoring programme. NIWA Consultancy Report.

Cummings, V.J., Thrush, S.F., Hewitt, J.E., Turner, S.J., 1998. The influence of the pinnid bivalve Atrina zelandica (Gray) on benthic macroinvertebrate communities in soft-sediment habitats. J. Exp. Mar. Biol. Ecol. 228, 227–240. doi:10.1016/S0022-0981(98)00028-8

Dayton, P.K., Tegner, M.J., Edwards, P.B., Riser, K.L., 1998. Sliding baselines, ghosts, and reduced expectations in kelp forest communities. Ecol. Appl. 8, 309–322.

de Juan, S., Demestre, M., 2012. A Trawl Disturbance Indicator to quantify large scale fishing impact on benthic ecosystems. Ecol. Ind. 18, 183–190. doi:10.1016/j.ecolind.2011.11.020

de Juan, S., Demestre, M., Thrush, S.F., 2009. Defining ecological indicators of trawling disturbance when everywhere that can be fished is fished: A Mediterranean case study. Mar. Pol. 33, 472–478. doi:10.1016/j.marpol.2008.11.005

de Juan, S., Gelcich, S., Fernández, M., 2017. Integrating stakeholder perceptions and preferences on ecosystem services in the management of coastal areas. Ocean Coast. Manage. 136, e12–e12. doi:10.1016/j.ocecoaman.2016.11.019

de Juan, S., Hewitt, J.E., 2013. Spatial and temporal variability in species richness in a temperate intertidal community. Ecography 37, 183–190. doi:10.1111/j.1600-0587.2013.00048.x

de Juan, S., Lo Iacono, C., Demestre, M., 2013. Benthic habitat characterisation of soft-bottom continental shelves: Integration of acoustic surveys, benthic samples and trawling disturbance intensity. Estuar. Coast. Shelf Sci. 117, 199–209. doi:10.1016/j.ecss.2012.11.012

de Juan, S., Subida, M.D., Gelcich, S., Fernandez, M., 2018. Ecosystem health in coastal areas targeted by small-scale artisanal fisheries: Insights on monitoring and assessment. Ecol. Ind. 88, 361–371. doi:10.1016/j.ecolind.2018.01.054

de Juan, S., Thrush, S.F., Demestre, M., 2007. Functional changes as indicators of trawling disturbance on a benthic community located in a fishing ground (NW Mediterranean Sea). Mar. Ecol. Prog. Ser. 334, 117–129.

Duarte, C.M., Conley, D.J., Carstensen, J., Sánchez-Camacho, M., 2008. Return to Neverland: Shifting Baselines Affect Eutrophication Restoration Targets. Estuaries Coasts 32, 29–36. doi:10.1007/s12237-008-9111-2

Edgar, G.J., Bates, A.E., Bird, T.J., Jones, A.H., Kininmonth, S., Stuart-Smith, R.D., Webb, T.J., 2016. New Approaches to Marine Conservation Through the Scaling Up of Ecological Data. Annu. Rev. Mar. Sci. 8, 435–461. doi:10.1146/annurev-marine-122414-033921

Elliott, M., 2011. Marine science and management means tackling exogenic unmanaged pressures and endogenic managed pressures – A numbered guide. Mar. Pollut. Bull. 62, 651–655.

Elliott, M., Borja, A., McQuatters-Gollop, A., Mazik, K., Birchenough, S., Andersen, J.H., Painting, S., Peck, M., 2015. Force majeure: Will climate change affect our ability to attain Good Environmental Status for marine biodiversity? Mar. Pollut. Bull. 95, 7–27. doi:10.1016/j.marpolbul.2015.03.015

Elliott, M., Burdon, D., Atkins, J.P., Borja, A., Cormier, R., de Jonge, V.N., Turner, R.K., 2017a. “And DPSIR begat DAPSI(W)R(M)!” - A unifying framework for marine environmental management. Mar. Pollut. Bull. 118, 27–40. doi:10.1016/j.marpolbul.2017.03.049

Elliott, M., Snoeijs-Leijonmalm, P., Barnard, S., 2017b. “The dissemination diamond” and paradoxes of science-to-science and science-to-policy communication: Lessons from large marine research programmes. Mar. Pollut. Bull. 125, 1–3. doi:10.1016/j.marpolbul.2017.08.022

Ellis, J.I., Cummings, V., Hewitt, J.E., Thrush, S.F., Norkko, A., 2002. Determining effects of suspended sediment on condition of a suspension feeding bivalve (Atrina zelandica): results of a survey, a laboratory experiment and a field transplant experiment. J. Exp. Mar. Biol. Ecol. 267, 147–174. doi:10.1016/S0022-0981(01)00355-0

Ellis, J.I., Norkko, A., Thrush, S.F., 2000. Broad-scale disturbance of intertidal and shallow sublittoral soft-sediment habitats; effects on the benthic macrofauna. Journal of Aquatic Ecosystem Stress and recovery 7, 57–74.

Estévez, R.A., Gelcich, S., 2015. Participative multi-criteria decision analysis in marine management and conservation: Research progress and the challenge of integrating value judgments and uncertainty. Mar. Pol. 61, 1–7. doi:10.1016/j.marpol.2015.06.022

European Commision, 2008. Directive 2008/56/EC of the European Parliament and of the Council establishing a framework for community action in the field of marine environmental policy (Marine Strategy Framework Directive).

European Commision, 2000. Directive 2000/60/EC of the European Parliament and of the Council of 23 Octobre 2000 establishing a framework for community action in the field of water policy.

Folke, C., Carpenter, S.R., Walker, B., Scheffer, M., Elmqvist, T., Gunderson, L., Holling, C.S., 2004. Regime shifts, resilience, and biodiversity in ecosystem management. Annu. Rev. Ecol. Evol. S. 35, 557–581. doi:10.1146/annurev.ecolsys.35.021103.105711

Game, E.T., Bremer, L.L., Calvache, A., Moreno, P.H., Vargas, A., Rivera, B., Rodriguez, L.M., 2017. Fuzzy Models to Inform Social and Environmental Indicator Selection for Conservation Impact Monitoring. Conserv. Let. 65, 579–9. doi:10.1111/conl.12338

GESAMP, 1994. Group of experts on the scientific aspects of marine environmental protection: Anthropogenic influences on sediment discharge to the coastal zone and environmental consequences. UNESCO-TOC, Paris.

Gray, J.S., 1977. The stability of benthic ecosystems. Helgoländer wissenschaftliche Meeresuntersuchungen 1977 30:1 30, 427–444. doi:10.1007/BF02207852

Halliday, J., Cummings, V.J., 2011. Mahurangi Estuary Ecological Monitoring Program: Report on Data Collected from July 1994 to January 2011.

Halpern, B.S., Longo, C., Hardy, D., McLeod, K.L., Samhouri, J.F., Katona, S.K., Kleisner, K., Lester, S.E., O'Leary, J., Ranelletti, M., Rosenberg, A.A., Scarborough, C., Selig, E.R., Best, B.D., Brumbaugh, D.R., Chapin, F.S., Crowder, L.B., Daly, K.L., Doney, S.C., Elfes, C., Fogarty, M.J., Gaines, S.D., Jacobsen, K.I., Karrer, L.B., Leslie, H.M., Neeley, E., Pauly, D., Polasky, S., Ris, B., St Martin, K., Stone, G.S., Sumaila, R., Zeller, D., 2012. An index to assess the health and benefits of the global ocean. Nature 488, 615–620. doi:10.1038/nature11397

Harding, L.E., 1992. Measures of marine environmental quality. Mar. Pollut. Bull. 25, 23–27. doi:10.1016/0025-326X(92)90178-9

Hayes, K.R., Dambacher, J.M., Hosack, G.R., Bax, N.J., Dunstan, P.K., Fulton, E.A., Thompson, P.A., Hartog, J.R., Hobday, A.J., Bradford, R., Foster, S.D., Hedge, P., Smith, D.C., Marshall, C.J., 2015. Identifying indicators and essential variables for marine ecosystems. Ecol. Ind. 57, 409–419. doi:10.1016/j.ecolind.2015.05.006

Hewitt, J.E., Anderson, M.J., Thrush, S.F., 2005. Assessing and monitoring ecological community health in marine systems. Ecol. Appl. 15, 942–953.

Hewitt, J.E., Cummings, V.J., Ellis, J.I., Funnell, G., Norkko, A., Talley, T.S., Thrush, S.F., 2003. The role of waves in the colonisation of terrestrial sediments deposited in the marine environment. J. Exp. Mar. Biol. Ecol. 290, 19–47. doi:10.1016/S0022-0981(03)00051-0

Hume, T.M., Morrisey, D.J., Elliot, S., 2009. A knowledge nework to inform sediment management: Mahurangi proof of concept Phase 2.

Jennings, S.R., Freeman, S.M., Parker, R.L., Duplisea, D.E., Dinmore, T.A., 2005. Ecosystem Consequences of Bottom Fishing Disturbance. American Fisheries Society, Florida (USA).

Kaiser, M.J., Clarke, K.R., Hinz, H., Austen, M.C., Somerfield, P.J., Karakassis, I., 2006. Global analysis of response and recovery of benthic biota to fishing. Mar. Ecol. Prog. Ser. 311, 1–14.

Kappel, C.V., Halpern, B.S., Martone, R., Micheli, F., Selkoe, K.A., 2009. In the Zone: Comprehensive Ocean Protection | Issues in Science and Technology. Issues Sci Technol 25.

Karr, J.R., 1993. Defining and assessing ecological integrity: Beyond water quality. Environmental Toxicology and Chemistry 12, 1521–1531. doi:10.1002/etc.5620120902

Kittinger, J.N., McClenachan, L., Gedan, K.B., Blight, L.K., 2015. Marine Historical Ecology in Conservation. Univ of California Press.

Konar, B., Estes, J.A., 2003. The stability of boundary regions between kelp beds and deforested areas. Ecology 84, 174–185. doi:10.1890/0012-9658(2003)084$[0174:TSOBRB]$2.0.CO;2

Ladle, R.J., Gillson, L., 2008. The (im)balance of nature: a public perception time-lag? Public Underst. Sci. 18, 229–242. doi:10.1177/0963662507082893

Lamy, T., Reed, D.C., Rassweiler, A., Siegel, D.A., Kui, L., Bell, T.W., Simons, R.D., Miller, R.J., 2018. Scale-specific drivers of kelp forest communities. Oecologia 186, 217–233. doi:10.1007/s00442-017-3994-1

Leslie, H.M., McLeod, K.L., 2007. Confronting the challenges of implementing marine ecosystem-based management. Front. Ecol. Environ. 5, 540–548. doi:10.1890/060093

Levin, P.S., Möllmann, C., 2014. Marine ecosystem regime shifts: challenges and opportunities for ecosystem-based management. Philosophical Transactions of the Royal Society B: Biological Sciences 370, 20130275–20130275. doi:10.1098/rstb.2013.0275

Lohrer, A.M., Hewitt, J.E., Thrush, S.F., 2006. Assessing far-field effects of terrigenous sediment loading in the coastal marine environment. Mar. Ecol. Prog. Ser. 315, 13–18. doi:10.3354/meps315013

Lonsdale, J.-A., Weston, K., Barnard, S., Boyes, S.J., Elliott, M., 2015. Integrating management tools and concepts to develop an estuarine planning support system: A case study of the Humber Estuary, Eastern England. Mar. Pollut. Bull. 100, 393–405. doi:10.1016/j.marpolbul.2015.08.017

Maani, K., Cavana, R.Y., 2007. Systems thinking, system dynamics: Managing change and complexity. Prentice Hall.

Mac Monagail, M., Cornish, L., Morrison, L., Araújo, R., Critchley, A.T., 2017. Sustainable harvesting of wild seaweed resources. European Journal of Phycology 52, 371–390. doi:10.1080/09670262.2017.1365273

Mangi, S.C., Roberts, C.M., Rodwell, L.D., 2007. Reef fisheries management in Kenya: Preliminary approach using the driver–pressure–state–impacts–response (DPSIR) scheme of indicators. Ocean Coast. Manage. 50, 463–480. doi:10.1016/j.ocecoaman.2006.10.003

Martínez-Crego, B., Alcoverro, T., Romero, J., 2010. Biotic indices for assessing the status of coastal waters: a review of strengths and weaknesses. J. Environ. Monit. 12, 1013–17. doi:10.1039/b920937a

McClenachan, L., Ferretti, F., \& Baum, J. K., 2012. From archives to conservation: why historical data are needed to set baselines for marine animals and ecosystems. Conserv. Lett., 5(5), 349-359.

MEA, 2005. Ecosystems and human well-being: synthesis. Washington DC.

Muntadas, A., de Juan, S., Demestre, M., 2015. Integrating the provision of ecosystem services and trawl fisheries for the management of the marine environment. Sci. Total Environ. 594–603. doi:10.1016/j.scitotenv.2014.11.042

Muntadas, A., Demestre, M., de Juan, S., Frid, C.L.J., 2014. Trawling disturbance on benthic ecosystems and consequences on commercial species: a northwestern Mediterranean case study. Sci. Mar. 78, 53–65. doi:10.3989/scimar.04024.19A

Muntadas, A., Lample, M., Demestre, M., Ballé-Béganton, J., de Juan, S., Maynou, F.X., Bailly, D., 2017. A knowledge platform to inform on the effects of trawling on benthic communities. Estuar. Coast. Shelf Sci. 1–11. doi:10.1016/j.ecss.2017.01.001

Müller, F., Burkhard, B., 2007. An ecosystem based framework to link landscape structures, functions and services. Springer, Berlin Heidelberg.

Nguyen, N.C., Bosch, O.J.H., 2012. A Systems Thinking Approach to identify Leverage Points for Sustainability: A Case Study in the Cat Ba Biosphere Reserve, Vietnam. Systems Research and Behavioral Science 30, 104–115. doi:10.1002/sres.2145

Norkko, A., Thrush, S.F., Hewitt, J.E., Cummings, V.J., Norkko, J., Ellis, J.I., Funnell, G.A., Schultz, D., MacDonald, I., 2002. Smothering of estuarine sandflats by terrigenous clay: the role of wind-wave disturbance and bioturbation in site-dependent macrofaunal recovery. Mar. Ecol. Prog. Ser. 234, 23–42. doi:10.3354/meps234023

Palanques, A., Guillen, J., Puig, P., 2001. Impact of bottom trawling on water turbidity and muddy sediment of an unfished continental shelf. Limnol. Oceanogr. 46, 1100–1110.

Paoli, C., Morten, A., Bianchi, C.N., Morri, C., Fabiano, M., Vassallo, P., 2016. Capturing ecological complexity: OCI, a novel combination of ecological indices as applied to benthic marine habitats. Ecol. Ind. 66, 86–102. doi:10.1016/j.ecolind.2016.01.029

Pereira, M., Tala, F., Fern ndez, M., Subida, M.D., 2015. Effects of kelp phenolic compounds on the feeding-associated mobility of the herbivore snail Tegula tridentata. Mar. Environ. Res. 112, 40–47. doi:10.1016/j.marenvres.2015.04.012

Pérez-Matus, A., Carrasco, S.A., Gelcich, S., Fernandez, M., Wieters, E.A., 2017a. Exploring the effects of fishing pressure and upwelling intensity over subtidal kelp forest communities in Central Chile. Ecosphere 8, e01808. doi:10.1002/ecs2.1808

Pérez-Matus, A., Ospina-Alvarez, A., Camus, P.A., Carrasco, S.A., Fernandez, M., Gelcich, S., Godoy, N., Ojeda, F.P., Pardo, L.M., Rozbaczylo, N., Subida, M.D., Thiel, M., Wieters, E.A., Navarrete, S.A., 2017b. Temperate rocky subtidal reef community reveals human impacts across the entire food web. Mar. Ecol. Prog. Ser. 567, 1–16. doi:10.3354/meps12057

Pinto, R., Patricio, J., Baeta, A., Fath, B.D., Neto, J.M., Marques, J.C., 2009. Review and evaluation of estuarine biotic indices to assess benthic condition. Ecol. Ind. 9, 1–25. doi:10.1016/j.ecolind.2008.01.005

Plumeridge, A.A., Roberts, C.M., 2017. Conservation targets in marine protected area management suffer from shifting baseline syndrome: A case study on the Dogger Bank. Mar. Pollut. Bull. 116, 395–404. doi:10.1016/j.marpolbul.2017.01.012

Rodil, I.F., Lohrer, A.M., Hewitt, J.E., Townsend, M., Thrush, S.F., Carbines, M., 2013. Tracking environmental stress gradients using three biotic integrity indices: Advantages of a locally-developed traits-based approach. Ecol. Ind. 34, 560–570. doi:10.1016/j.ecolind.2013.06.023

Roelfsema, C., Thurstan, R., Beger, M., Dudgeon, C., Loder, J., Kovacs, E., Gallo, M., Flower, J., Gomez Cabrera, K.-L., Ortiz, J., Lea, A., Kleine, D., 2016. A Citizen Science Approach: A Detailed Ecological Assessment of Subtropical Reefs at Point Lookout, Australia. PloS ONE 11, e0163407–20. doi:10.1371/journal.pone.0163407

Roemmich, D., Gould, W.J., Gilson, J., 2012. 135 years of global ocean warming between the Challenger expedition and the Argo Program. Nature Climate Change 2, nclimate1461–428. doi:10.1038/nclimate1461

Rombouts, I., Beaugrand, G., Artigas, L.F., Dauvin, J.C., Gevaert, F., Goberville, E., Kopp, D., Lefebvre, S., Luczak, C., Spilmont, N., Travers-Trolet, M., Villanueva, M.C., Kirby, R.R., 2013. Evaluating marine ecosystem health: Case studies of indicators using direct observations and modelling methods. Ecol. Ind. 24, 353–365. doi:10.1016/j.ecolind.2012.07.001

Ruano-Chamorro, C., Subida, M.D., Fernandez, M., 2017. Fishers' perception: An alternative source of information to assess the data-poor benthic small-scale artisanal fisheries of central Chile. Ocean Coast. Manage. 146, 67–76. doi:10.1016/j.ocecoaman.2017.06.007

Santelices, B., Ojeda, F.P., 1984. Recruitment, growth and survival of Lessonia nigrescens (Phaeophyta) at various tidal levels in exposed habitats of central Chile. Mar. Ecol. Prog. Ser. 19, 73–82. doi:10.3354/meps019073

Scheffer, M., Bascompte, J., Brock, W.A., Brovkin, V., Carpenter, S.R., Dakos, V., Held, H., van Nes, E.H., Rietkerk, M., Sugihara, G., 2009. Early-warning signals for critical transitions. Nature 461, 53–59. doi:10.1038/nature08227

Scheffer, M., Carpenter, S.R., Foley, J.A., Folke, C., Walker, B., 2001. Catastrophic shifts in ecosystems. Nature 413, 591–596.

Snelgrove, P.V.R., Butman, C.A., 1995. Animal-sediment relationships revisited: cause versus effect. Oceanographic Literature Review 8, 668.

Tett, P., Gowen, R.J., Painting, S.J., Elliot, M., Forster, R., Mills, D.K., Bresnan, E., Capuzzo, E., Fernandes, T.F., Foden, J., Geider, R.J., Gilpin, L.C., Huxham, M., McQuatters-Gollop, A.L., Malcolm, S.J., Saux-Picart, S., Platt, T., Racault, M.-F., Sathyendranath, S., van der Molen, J., Wilkinson, M., 2013. Framework for understanding marine ecosystem health. Mar. Ecol. Prog. Ser. 494, 1–27.

Thrush, S.F., Halliday, J., Hewitt, J.E., Lohrer, A.M., 2008a. The effects of habitat loss, fragmentation, and community homogenization on resilience in estuaries. Ecol. Appl. 18, 12–21.

Thrush, S.F., Hewitt, J.E., Dayton, P.K., Coco, G., Lohrer, A.M., Norkko, A., Norkko, J., Chiantore, M., 2009. Forecasting the limits of resilience: integrating empirical research with theory. Proc. Biol. Sci. 276, 3209–3217. doi:10.1098/rspb.2009.0661

Thrush, S.F., Hewitt, J.E., Hickey, C.W., Kelly, S., 2008b. Multiple stressor effects identified from species abundance distributions: Interactions between urban contaminants and species habitat relationships. J. Exp. Mar. Biol. Ecol. 366, 160–168. doi:10.1016/j.jembe.2008.07.020

Thrush, S.F., Hewitt, J.E., Lohrer, A.M., 2012. Interaction networks in coastal soft-sediments highlight the potential for change in ecological resilience. Ecol. Appl. 22, 1213–1223.

Thrush, S.F., Hewitt, J.E., Lohrer, A.M., Chiaroni, L.D., 2013. When small changes matter: the role of cross-scale interactions between habitat and ecological connectivity in recovery. Ecol. Appl. 23, 226–238. doi:10.1890/12-0793.1

Thrush, S.F., Hewitt, J.E., Norkko, A., Cummings, V.J., Funnell, G.A., 2003. Macrobenthic recovery processes following catastrophic sedimentation on estuarine sandflats. Ecol. Appl. 13, 1433–1455.

Van Hoey, G., Borja, A., Birchenough, S., Buhl-Mortensen, L., Degraer, S., Fleischer, D., Kerckhof, F., Magni, P., Muxika, I., Reiss, H., Schröder, A., Zettler, M., 2010. The use of benthic indicators in Europe: from the Water Framework Directive to the Marine Strategy Framework Directive. Mar. Pollut. Bull. 60, 2187–2196. doi:10.1016/j.marpolbul.2010.09.015

Varner, J., 2014. Scientific Outreach: Toward Effective Public Engagement with Biological Science. BioScience 64, 333–340. doi:10.1093/biosci/biu021

Vásquez, J.A., Piaget, N., Vega, J.M.A., 2012. The Lessonia nigrescens fishery in northern Chile: “how you harvest is more important than how much you harvest.” J. Appl. Phycol. 24, 417–426. doi:10.1007/s10811-012-9794-4

Vásquez, J.A., Santelices, B., 1990. Ecological effects of harvesting Lessonia (Laminariales, Phaeophyta) in central Chile. Hydrobiologia 204-205, 41–47. doi:10.1007/BF00040213

\end{document}